\documentstyle[sprocl,epsf]{article}
\def\be{\begin{equation}}
\def\ee{\end{equation}}
\def\bea{\begin{eqnarray}}
\def\eea{\end{eqnarray}}

\def\hide#1{}

\let\ga=\gamma

\let\De=\Delta

\let\del=\nabla
\let\si=\sigma

\let\Om=\Omega

\def\seq {\! = \!} 
\def\nl{\hfil\break}

\let\<=\langle
\let\>=\rangle

\let\txt=\textstyle

\def\beq{\begin{equation}}
\def\eeq{\end{equation}}
\def\ba{\begin{array}}
\def\bea{\begin{eqnarray}}
\def\ea{\end{array}}
\def\eea{\end{eqnarray}}
\newcommand{\nn}{\nonumber \\}

\def\slash{\!\!\!\!/\,}

\def\comment#1{ \hbox{[{\it Comment suppressed here.}\/]} }
\def\o{\over}      
\def\O{ {\cal O} }

\def\half {{\txt {1\over 2}}}

\def\Fmn{F_{\mu\nu}}

\def\psib{{\bar\psi}}
\def\Omb{{\bar\Om}}

\def\={\!=\!}
\def\+{\,+\,}
\def\-{\,-\,}

\def\sigF{\si \! \cdot \! F}

\begin{document}

\title{IMPROVED QUARK ACTIONS FOR LATTICE QCD}

\author{Timothy R. Klassen}

\address{SCRI, Florida State University, Tallahassee, FL 32306-4052, USA}

\maketitle\abstracts{
I present a brief summary of the status and prospects of improved
Wilson-type quark actions for coarse lattice simulations. My
conclusions are optimistic.
}

\section{Introduction}\label{sec:intro}

In the last few years it has become clear that there are a variety of
gluon actions that give accurate results on coarse 
lattices ($a\seq 0.2 - 0.4$~fm); see various contributions to 
Lattice~96.\cite{lat96} Such actions
have been constructed within the Symanzik improvement program,
eliminating $\O(a^2)$ errors using tadpole improvement~\cite{TI} at 
tree- and at one-loop level, 
as well as within the MCRG (or ``perfect action'') approach. 
 There are of course several
open problems that should be addressed,\footnote{For example, finding
a practical method of non-perturbatively eliminating {\it all} $\O(a^2)$
errors of Symanzik improved gluon actions 
(not just the violation of rotational symmetry),
or, whether the second-order phase
transition in the fundamental-adjoint plane of gauge couplings affects
the physics of these actions on coarse lattices, in particular the
$0^{++}$ glueball.}  
but, generally speaking, the errors of these actions on coarse
lattices seem quite small, much smaller than those of any improved quark 
action proposed so far. Given the dramatic increase in cost of a full
QCD simulation as the lattice spacing is decreased,
it is very
important to find improved quark actions that are accurate on coarse
lattices. This is the aim I address in this contribution.

Besides the use of improved actions, another tool that seems likely to
become a staple of lattice QCD technology, is the use of {\it anisotropic
lattices},\cite{Kar,Stam,ILQA,aniso}
 with smaller temporal than spatial lattice spacing,
$a_0 \equiv a_t < a_s \equiv a_i ~(i=1,2,3)$. 
[A lattice with $\xi = a_s/a_t$ will be referred to as a
``$\xi:1$ lattice''.]
Such lattices have clear
advantages in the study of heavy quarks,\footnote{Important, since
NRQCD methods  seem to   break down for charmonium.~\cite{Trot}}
lattice thermodynamics,
and for particles with bad signal/noise properties, like glueballs
and P-state mesons. On the classical level anisotropic lattices are
as easy to treat as isotropic ones. On the quantum level, however,
more coefficients have to be tuned to restore space-time exchange
symmetry. Perturbative calculations~\cite{vBaniso} and preliminary simulations
with heavy quarks~\cite{LAT96} and glueballs~\cite{MorPea} have appeared using
improved anisotropic gluon actions;
further work is in progress.\cite{Stam,aniso}

\section{Doubler-Free, Classically Improved Quark Actions}

The first step in the Symanzik improvement program is the construction
of a {\it classically} improved action. Usually this is quite simple;
for fermions one has a slight complication due to the doubler problem.
Recall that the origin of doublers is that the standard discretization
of the  continuum derivative $D_\mu$,
\beq
 \del_\mu \psi(x) ~\equiv~
 {1\over 2a_\mu}\, \biggl[ U_\mu(x) \psi(x+\mu) - 
                       U_{-\mu}(x) \psi(x-\mu)\biggr] 
 ~=~ D_\mu \psi(x) + \O(a_\mu^2) ~,
\eeq
{\it decouples even and odd} sites of the lattice. Wilson suggested to avoid 
this problem by adding  a second-order derivative term 
$\psib \sum_\mu \De_\mu \, \psi$ to 
the action, 
\beq
\De_\mu \psi(x)  \, \equiv \, {1\o a_\mu^2}
     \biggl[U_\mu(x) \psi(x+\mu) + U_{-\mu}(x) \psi(x-\mu) -2\psi(x) \biggr] 
    = D^2_\mu \psi(x) + \O(a_\mu^2).
\eeq
If one adds such a term naively, one breaks chiral symmetry at $\O(a)$.
But chiral symmetry can be preserved to higher order if one introduces the
{\it Wilson term} $\sum_\mu \De_\mu$ by a {\it field transformation}. 
The simplest way to proceed is to start with the continuum action 
$\psib_c \, M_c \, \psi_c \equiv \psib (D\slash + m_c)\psi$ and perform
the field redefinition
$\psi_c = \Om_c \, \psi, ~\psib_c =\psib \, \Omb_c$ with~\footnote{The
Jacobian of a field transformation matters only at the quantum 
level, where, in the case at hand, its leading effect is to
renormalize the gauge coupling.}
\beq\label{cvstd}
\Omb_c = \Om_c ~,  \quad 
 \Omb_c \, \Om_c = 1 \- {r a_0 \over 2} \, (D\slash - m_c) ~.
\eeq
Here $r$ is a free parameter, to be fixed later. The transformed fermion
operator reads
\bea\label{MOm}
 \Omb_c \, M_c \, \Om_c ~=~ m_c (1 + \half r a_0 m_c) \+ D\slash
       - {1\over 2} \, r a_0 \,
   \Bigl(\sum_\mu D_\mu^2 \+ \half \sigF \Bigr) ~,
\eea
where we used $D\slash^2 = \sum_\mu D^2_\mu + \half \sigF$.
Here $\sigF \equiv \sum_{\mu\nu} \si_{\mu\nu} \Fmn$
is the {\it clover term}, containing
the field strength $\Fmn$. The above continuum action still has
(slightly hidden) chiral symmetry. We can {\it now} discretize this
action by replacing $D\slash, D_\mu^2$ and $\Fmn$ by suitable lattice
versions, differing at $\O(a^n)$, say,  from the former.
Let us call the action so obtained  $M$. It 
will not have a doubler problem. 
It will break chiral symmetry, however; classically at $\O(a_0 a^n)$, 
on the quantum level
at $\O(a_0 g^2)$. $M$ will correctly give {\it on-shell} quantities
up to $\O(a^n)$ errors at tree level. Off-shell quantities can also be 
obtained with such errors, by simply {\it undoing} the field transformation
on the lattice.

For $n\seq 4$ one obtains the D234 actions~\cite{ILQA} on an anisotropic 
lattice,
\bea\label{MDiiii}
 M_{{\rm D234}} &=&  m_c (1 + \half r a_0 m_c) \+ \sum_\mu\,
      \ga_\mu \del_\mu \, ( 1 - b_\mu a_\mu^2 \De_\mu ) \nn
&& \- {1\over 2} \, r a_0  \,
    \Bigl( \sum_\mu \Delta_\mu  \, \+ \, \half \, \si \! \cdot \! F \Bigr)
            \+ \sum_\mu c_\mu a_\mu^3 \De_\mu^2 
\eea
where $b_\mu = {1\o 6}$ and  $c_\mu = {r a_0\over 24 a_\mu}$.
The SW (or clover) action~\cite{SW} is the $n\seq 2$ case, corresponding to 
$b_\mu \seq c_\mu \seq 0$. The Wilson action is obtained from the SW
action by ignoring the clover term $\sigF$.

A generic property of improved actions is the existence of unphysical
branches in their (free) dispersion relations, arising from higher
order temporal derivatives in the action. We will refer to such branches
as {\it ghost} branches.  The SW action has no ghost branches for $r\seq 1$,
which is therefore the canonical choice in this case. In general one can
not eliminate all ghost branches. The D234 action as derived above with
$\O(a^4)$ errors has three ghost branches (except for $r\seq 2$, when
there are two). It is not clear if this is really a problem, but we have
first considered slightly modified D234 actions that have only one ghost
branch, at the expense of introducing small $a^3$ or $a_0^3$ errors
for isotropic, respectively, anisotropic lattices. The former case,
the ``isotropic D234'' action~\cite{LAT95} corresponds to choosing 
$r\seq {2\o 3}$
and $c_\mu \seq {1\o 12}$. The latter, the anisotropic ``D234i(${2\o 3}$)''
action, has $r\seq {2\o 3}$, $c_0\seq {1\o 12}$ 
(and $c_i = {r a_0\over 24 a_i}$ as originally derived). 

For more details about improved quark actions, including plots of
various dispersion relations, we refer the reader to ref.\cite{ILQA}

\section{On-Shell Quantum Improvement}

For both gluon and quark actions the largest error at $\O(a^2)$ is the
violation of rotational symmetry, which already exists at the classical
level. In fact, in the case of gluons no significant differences have
so far been found~\cite{Alf1} between actions that correct only the 
violation of rotational symmetry and others that also take into account 
$a^2$ terms that are only necessary on the quantum level. Whereas there
is some hope that all (two) terms necessary for the $\O(a^2)$ on-shell
improvement of a gluonic action can eventually be tuned non-perturbatively,
this is clearly impossible for a Wilson-type quark action; there are just
too many terms.\cite{SW} But it is not unreasonable to hope that, as for glue,
the largest $a^2$ errors of quark actions are the violations of rotational
symmetry.

A clear difference between gluon and Wilson-type quark actions 
emerges~at $\O(a)$. No such terms are present for gluons, but for quarks we 
have to introduce Wilson and clover terms to eliminate doublers without 
introducing classical $\O(a)$ errors. In this case these two terms are also 
the only ones that can exist at the quantum level at this order. The
coefficient of one of these terms, usually chosen to be the Wilson term,
can be adjusted at will by a field redefinition. The other one then has
to be tuned to eliminate $\O(a)$ quantum errors.

 Note that the $\O(a)$
terms break chiral but not rotational symmetry; exactly the opposite
behavior of the leading $a^2$ terms. This qualitative difference allows
one to tune both, even non-perturbatively, by demanding the restoration
of chiral symmetry at zero quark mass for the former 
(how to implement this in practice has recently been shown in important
work by L\"uscher et~al~\cite{LPCAC}),
that of rotational symmetry for the latter.

So far we have discussed isotropic lattices. The anisotropic case is more
complicated. However, after considering the most general field redefinitions
up to $\O(a)$, one easily sees that only two more parameters have to be
tuned for on-shell improvement of a quark action up to $\O(a)$. One
already appears at $\O(a^0)$, namely, a ``bare velocity of light''
that has to be tuned to restore space-time exchange symmetry (by, say,
demanding that the pion have a relativistic dispersion relation for small
masses and momenta). The other is at $\O(a)$; the two terms that now have
to be tuned at this order can be chosen to be the temporal and spatial
parts of the clover term.

\section{Quenched Simulation Results with Tadpole Improved Actions}

We would now like to use simulation results for various tadpole improved
actions in an attempt to disentangle the effects of the $\O(a)$ and 
$\O(a^2)$ terms, as a handle on how well tadpole improvement (TI) 
estimates the coefficients of these terms.
To alleviate the problem of (absolute) scale setting, we will 
concentrate on dimensionless quantities and compare results obtained
with the Wilson, SW and D234 actions with the same improved gluon
actions. Specifically, we consider:\nl
(a) The ``effective velocity of
light'',  $c({\bf p})$, of various hadrons, defined by
$c({\bf p})^2 {\bf p}^2 = E({\bf p})^2 - E(0)^2$.\nl
(b) $J$, a dimensionless measure of the vector versus pseudo-scalar meson
mass relation, defined~\cite{J} by
$J=m_V dm_V/dm_P^2$ ~at~ $m_V/m_P = m_{K^\star}/m_K = 1.8$.\footnote{As
an aside we remark that $J$ is one of the most accurately known numbers in
quenched continuum QCD, 
$J=0.39(1)$ (see below).
It also seems to be
a very sensitive indicator of quenching errors, since in nature
$J=0.48(2)$.\cite{J}}\nl
(c) The nucleon over rho mass ratio, $m_N/m_\rho$ (defined by extrapolation
of lattice data to $m_\rho/m_\pi = 5.58$).\nl
(d) The quenched hyper-fine splitting (HFS) of charmonium.\nl
(e) The rho mass in units of the string tension, $m_\rho/\sqrt{\si}$.\nl
Obviously, $c({\bf p})$ is a measure of $\O(a^2)$ violations of rotational
symmetry, and therefore essentially independent of the clover coefficient.
The HFS and the rho mass, on the other hand, clearly depend strongly on the
clover coefficient. They also have some dependence on the $\O(a^2)$
terms (see below). For $J$ and $m_N/m_\rho$
it seems~\cite{UKQCD,Gock,SCRI} (see also table~2) that they have a 
significant dependence on both $\O(a)$ and $\O(a^2)$
terms, though their dependence on the clover 
coefficient is much weaker than that of $m_\rho/\sqrt{\si}$ (certainly 
on finer lattices).

\begin{table}[bt]
\caption{
$c({\bf p})^2$ for mesons with momentum $p \seq 2\pi/aL$, 
$aL\! \approx \! 2.0$~fm at $m_\rho/m_\pi \approx 1.4$,
calculated for various actions on tadpole and one-loop improved 
quenched glue.
}
\label{tab:csq}
\vspace{0.3cm}
\begin{center}
\begin{tabular}{| l | ll | ll | ll |}
\hline
&   \makebox[1mm][l]{D234~\protect\cite{LAT95}} 
&&  \makebox[1mm][l]{D234 (no TI)}
&&  \makebox[1mm][l]{SW~\protect\cite{LAT95}} & \\
$\beta$ & $\pi$ & $\rho$ & $\pi$ & $\rho$ & $\pi$ & $\rho$ \\
\hline
6.8  &  0.95(2)  & 0.93(3)  & 0.83(2) & 0.75(4)  & 0.63(2) & 0.48(3) \\
7.1  &  0.94(3)  & 0.96(5)  & ---     & ---      & 0.74(3) & 0.55(4) \\
7.4  &  0.99(4)  & 1.00(6)  & ---     & ---      & ---     & --- \\
\hline
\end{tabular}
\end{center}
\end{table}

\begin{table}[bt]
\setlength{\tabcolsep}{0.4pc}     
\caption{
$J$ and $m_N/m_\rho$ for various tadpole improved actions.
The quenched continuum 
limit of $m_N/m_\rho \seq 1.29(2)$.\protect\cite{Urs}
The data~\protect\cite{LAT95} were partially reanalyzed.
}
\label{tab:Jetc}
\vspace{0.2cm}
\begin{center}
\begin{tabular}{| l | l | ll | ll | ll |}
\hline
& & \makebox[1mm][l]{D234~\protect\cite{LAT95}} 
&&  \makebox[1mm][l]{SW~\protect\cite{SCRI}}
&&  \makebox[1mm][l]{Wilson~\protect\cite{SCRI}} & \\
$\beta$ &$a$(fm) &$J$ &$m_N/m_\rho$ &$J$ &$m_N/m_\rho$&$J$ &$m_N/m_\rho$\\
\hline
6.8  & 0.40 & 0.386(5) & 1.40(4) & 0.345(4) & 1.46(2) & 0.314(3) & 1.78(1) \\
7.1  & 0.33 & 0.381(6) & 1.26(4) & 0.350(4) & 1.36(1) & 0.318(3) & 1.60(2) \\
7.4  & 0.24 & 0.395(10)& 1.27(8) & 0.371(5) & 1.34(2) & 0.335(5) & 1.56(3) \\
7.75 & 0.18 & ---      & ---     & 0.386(9) & 1.31(3) & 0.350(6) & 1.41(3) \\
\hline
\end{tabular}
\end{center}
\end{table}

\def\figHFS{
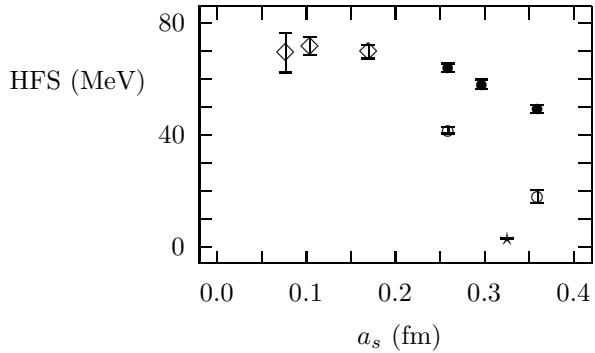
\begin{figure}[htb]  
\begin{center}
\setlength{\unitlength}{0.240900pt}
\ifx\plotpoint\undefined\newsavebox{\plotpoint}\fi
\begin{picture}(900,540)(0,0)
\font\gnuplot=cmr10 at 10pt
\gnuplot
\sbox{\plotpoint}{\rule[-0.200pt]{0.400pt}{0.400pt}}%
\put(220.0,139.0){\rule[-0.200pt]{4.818pt}{0.400pt}}
\put(198,139){\makebox(0,0)[r]{0}}
\put(816.0,139.0){\rule[-0.200pt]{4.818pt}{0.400pt}}
\put(220.0,183.0){\rule[-0.200pt]{4.818pt}{0.400pt}}
\put(816.0,183.0){\rule[-0.200pt]{4.818pt}{0.400pt}}
\put(220.0,227.0){\rule[-0.200pt]{4.818pt}{0.400pt}}
\put(816.0,227.0){\rule[-0.200pt]{4.818pt}{0.400pt}}
\put(220.0,271.0){\rule[-0.200pt]{4.818pt}{0.400pt}}
\put(816.0,271.0){\rule[-0.200pt]{4.818pt}{0.400pt}}
\put(220.0,315.0){\rule[-0.200pt]{4.818pt}{0.400pt}}
\put(198,315){\makebox(0,0)[r]{40}}
\put(816.0,315.0){\rule[-0.200pt]{4.818pt}{0.400pt}}
\put(220.0,359.0){\rule[-0.200pt]{4.818pt}{0.400pt}}
\put(816.0,359.0){\rule[-0.200pt]{4.818pt}{0.400pt}}
\put(220.0,403.0){\rule[-0.200pt]{4.818pt}{0.400pt}}
\put(816.0,403.0){\rule[-0.200pt]{4.818pt}{0.400pt}}
\put(220.0,447.0){\rule[-0.200pt]{4.818pt}{0.400pt}}
\put(816.0,447.0){\rule[-0.200pt]{4.818pt}{0.400pt}}
\put(220.0,491.0){\rule[-0.200pt]{4.818pt}{0.400pt}}
\put(198,491){\makebox(0,0)[r]{80}}
\put(816.0,491.0){\rule[-0.200pt]{4.818pt}{0.400pt}}
\put(248.0,113.0){\rule[-0.200pt]{0.400pt}{4.818pt}}
\put(248,68){\makebox(0,0){0.0}}
\put(248.0,497.0){\rule[-0.200pt]{0.400pt}{4.818pt}}
\put(318.0,113.0){\rule[-0.200pt]{0.400pt}{4.818pt}}
\put(318.0,497.0){\rule[-0.200pt]{0.400pt}{4.818pt}}
\put(388.0,113.0){\rule[-0.200pt]{0.400pt}{4.818pt}}
\put(388,68){\makebox(0,0){0.1}}
\put(388.0,497.0){\rule[-0.200pt]{0.400pt}{4.818pt}}
\put(458.0,113.0){\rule[-0.200pt]{0.400pt}{4.818pt}}
\put(458.0,497.0){\rule[-0.200pt]{0.400pt}{4.818pt}}
\put(528.0,113.0){\rule[-0.200pt]{0.400pt}{4.818pt}}
\put(528,68){\makebox(0,0){0.2}}
\put(528.0,497.0){\rule[-0.200pt]{0.400pt}{4.818pt}}
\put(598.0,113.0){\rule[-0.200pt]{0.400pt}{4.818pt}}
\put(598.0,497.0){\rule[-0.200pt]{0.400pt}{4.818pt}}
\put(668.0,113.0){\rule[-0.200pt]{0.400pt}{4.818pt}}
\put(668,68){\makebox(0,0){0.3}}
\put(668.0,497.0){\rule[-0.200pt]{0.400pt}{4.818pt}}
\put(738.0,113.0){\rule[-0.200pt]{0.400pt}{4.818pt}}
\put(738.0,497.0){\rule[-0.200pt]{0.400pt}{4.818pt}}
\put(808.0,113.0){\rule[-0.200pt]{0.400pt}{4.818pt}}
\put(808,68){\makebox(0,0){0.4}}
\put(808.0,497.0){\rule[-0.200pt]{0.400pt}{4.818pt}}
\put(220.0,113.0){\rule[-0.200pt]{148.394pt}{0.400pt}}
\put(836.0,113.0){\rule[-0.200pt]{0.400pt}{97.324pt}}
\put(220.0,517.0){\rule[-0.200pt]{148.394pt}{0.400pt}}
\put(23,315){\makebox(0,0){\raise 15mm\hbox{ HFS (MeV)}}}
\put(528,23){\makebox(0,0){\lower 7mm\hbox{{ $a_s ~({\rm fm})$}}}}
\put(220.0,113.0){\rule[-0.200pt]{0.400pt}{97.324pt}}
\put(611,420){\circle*{18}}
\put(664,394){\circle*{18}}
\put(751,356){\circle*{18}}
\put(611.0,414.0){\rule[-0.200pt]{0.400pt}{3.132pt}}
\put(601.0,414.0){\rule[-0.200pt]{4.818pt}{0.400pt}}
\put(601.0,427.0){\rule[-0.200pt]{4.818pt}{0.400pt}}
\put(664.0,387.0){\rule[-0.200pt]{0.400pt}{3.613pt}}
\put(654.0,387.0){\rule[-0.200pt]{4.818pt}{0.400pt}}
\put(654.0,402.0){\rule[-0.200pt]{4.818pt}{0.400pt}}
\put(751.0,349.0){\rule[-0.200pt]{0.400pt}{3.132pt}}
\put(741.0,349.0){\rule[-0.200pt]{4.818pt}{0.400pt}}
\put(741.0,362.0){\rule[-0.200pt]{4.818pt}{0.400pt}}
\put(611,322){\circle{18}}
\put(751,218){\circle{18}}
\put(611.0,317.0){\rule[-0.200pt]{0.400pt}{2.409pt}}
\put(601.0,317.0){\rule[-0.200pt]{4.818pt}{0.400pt}}
\put(601.0,327.0){\rule[-0.200pt]{4.818pt}{0.400pt}}
\put(751.0,208.0){\rule[-0.200pt]{0.400pt}{4.818pt}}
\put(741.0,208.0){\rule[-0.200pt]{4.818pt}{0.400pt}}
\put(741.0,228.0){\rule[-0.200pt]{4.818pt}{0.400pt}}
\put(704,152){\makebox(0,0){$\star$}}
\put(704.0,151.0){\usebox{\plotpoint}}
\put(694.0,151.0){\rule[-0.200pt]{4.818pt}{0.400pt}}
\put(694.0,152.0){\rule[-0.200pt]{4.818pt}{0.400pt}}
\put(356,444){\raisebox{-.8pt}{\makebox(0,0){$\Diamond$}}}
\put(394,454){\raisebox{-.8pt}{\makebox(0,0){$\Diamond$}}}
\put(486,446){\raisebox{-.8pt}{\makebox(0,0){$\Diamond$}}}
\put(356.0,413.0){\rule[-0.200pt]{0.400pt}{14.936pt}}
\put(346.0,413.0){\rule[-0.200pt]{4.818pt}{0.400pt}}
\put(346.0,475.0){\rule[-0.200pt]{4.818pt}{0.400pt}}
\put(394.0,440.0){\rule[-0.200pt]{0.400pt}{6.986pt}}
\put(384.0,440.0){\rule[-0.200pt]{4.818pt}{0.400pt}}
\put(384.0,469.0){\rule[-0.200pt]{4.818pt}{0.400pt}}
\put(486.0,435.0){\rule[-0.200pt]{0.400pt}{5.059pt}}
\put(476.0,435.0){\rule[-0.200pt]{4.818pt}{0.400pt}}
\put(476.0,456.0){\rule[-0.200pt]{4.818pt}{0.400pt}}
\end{picture}
\end{center}
\vspace{-3mm}
\caption{
Quenched charmonium hyper-fine splitting for various tadpole improved
actions:
D234(${2\o 3}$) 3:1 ($\bullet$), SW 3:1 ($\circ$),
SW 1:1 ($\star$), FNAL ($\diamond$)~\protect\cite{FNAL} (a not fully 
relativistic formalism).
}
\label{fig:HFS}
\end{figure}
}

\begin{figure}[htb]  
\begin{center}
\setlength{\unitlength}{0.240900pt}
\ifx\plotpoint\undefined\newsavebox{\plotpoint}\fi
\begin{picture}(900,540)(0,0)
\font\gnuplot=cmr10 at 10pt
\gnuplot
\sbox{\plotpoint}{\rule[-0.200pt]{0.400pt}{0.400pt}}%
\put(220.0,139.0){\rule[-0.200pt]{4.818pt}{0.400pt}}
\put(198,139){\makebox(0,0)[r]{0}}
\put(816.0,139.0){\rule[-0.200pt]{4.818pt}{0.400pt}}
\put(220.0,183.0){\rule[-0.200pt]{4.818pt}{0.400pt}}
\put(816.0,183.0){\rule[-0.200pt]{4.818pt}{0.400pt}}
\put(220.0,227.0){\rule[-0.200pt]{4.818pt}{0.400pt}}
\put(816.0,227.0){\rule[-0.200pt]{4.818pt}{0.400pt}}
\put(220.0,271.0){\rule[-0.200pt]{4.818pt}{0.400pt}}
\put(816.0,271.0){\rule[-0.200pt]{4.818pt}{0.400pt}}
\put(220.0,315.0){\rule[-0.200pt]{4.818pt}{0.400pt}}
\put(198,315){\makebox(0,0)[r]{40}}
\put(816.0,315.0){\rule[-0.200pt]{4.818pt}{0.400pt}}
\put(220.0,359.0){\rule[-0.200pt]{4.818pt}{0.400pt}}
\put(816.0,359.0){\rule[-0.200pt]{4.818pt}{0.400pt}}
\put(220.0,403.0){\rule[-0.200pt]{4.818pt}{0.400pt}}
\put(816.0,403.0){\rule[-0.200pt]{4.818pt}{0.400pt}}
\put(220.0,447.0){\rule[-0.200pt]{4.818pt}{0.400pt}}
\put(816.0,447.0){\rule[-0.200pt]{4.818pt}{0.400pt}}
\put(220.0,491.0){\rule[-0.200pt]{4.818pt}{0.400pt}}
\put(198,491){\makebox(0,0)[r]{80}}
\put(816.0,491.0){\rule[-0.200pt]{4.818pt}{0.400pt}}
\put(248.0,113.0){\rule[-0.200pt]{0.400pt}{4.818pt}}
\put(248,68){\makebox(0,0){0.0}}
\put(248.0,497.0){\rule[-0.200pt]{0.400pt}{4.818pt}}
\put(318.0,113.0){\rule[-0.200pt]{0.400pt}{4.818pt}}
\put(318.0,497.0){\rule[-0.200pt]{0.400pt}{4.818pt}}
\put(388.0,113.0){\rule[-0.200pt]{0.400pt}{4.818pt}}
\put(388,68){\makebox(0,0){0.1}}
\put(388.0,497.0){\rule[-0.200pt]{0.400pt}{4.818pt}}
\put(458.0,113.0){\rule[-0.200pt]{0.400pt}{4.818pt}}
\put(458.0,497.0){\rule[-0.200pt]{0.400pt}{4.818pt}}
\put(528.0,113.0){\rule[-0.200pt]{0.400pt}{4.818pt}}
\put(528,68){\makebox(0,0){0.2}}
\put(528.0,497.0){\rule[-0.200pt]{0.400pt}{4.818pt}}
\put(598.0,113.0){\rule[-0.200pt]{0.400pt}{4.818pt}}
\put(598.0,497.0){\rule[-0.200pt]{0.400pt}{4.818pt}}
\put(668.0,113.0){\rule[-0.200pt]{0.400pt}{4.818pt}}
\put(668,68){\makebox(0,0){0.3}}
\put(668.0,497.0){\rule[-0.200pt]{0.400pt}{4.818pt}}
\put(738.0,113.0){\rule[-0.200pt]{0.400pt}{4.818pt}}
\put(738.0,497.0){\rule[-0.200pt]{0.400pt}{4.818pt}}
\put(808.0,113.0){\rule[-0.200pt]{0.400pt}{4.818pt}}
\put(808,68){\makebox(0,0){0.4}}
\put(808.0,497.0){\rule[-0.200pt]{0.400pt}{4.818pt}}
\put(220.0,113.0){\rule[-0.200pt]{148.394pt}{0.400pt}}
\put(836.0,113.0){\rule[-0.200pt]{0.400pt}{97.324pt}}
\put(220.0,517.0){\rule[-0.200pt]{148.394pt}{0.400pt}}
\put(23,315){\makebox(0,0){\raise 15mm\hbox{ HFS (MeV)}}}
\put(528,23){\makebox(0,0){\lower 7mm\hbox{{ $a_s ~({\rm fm})$}}}}
\put(220.0,113.0){\rule[-0.200pt]{0.400pt}{97.324pt}}
\put(611,420){\circle*{18}}
\put(664,394){\circle*{18}}
\put(751,356){\circle*{18}}
\put(611.0,414.0){\rule[-0.200pt]{0.400pt}{3.132pt}}
\put(601.0,414.0){\rule[-0.200pt]{4.818pt}{0.400pt}}
\put(601.0,427.0){\rule[-0.200pt]{4.818pt}{0.400pt}}
\put(664.0,387.0){\rule[-0.200pt]{0.400pt}{3.613pt}}
\put(654.0,387.0){\rule[-0.200pt]{4.818pt}{0.400pt}}
\put(654.0,402.0){\rule[-0.200pt]{4.818pt}{0.400pt}}
\put(751.0,349.0){\rule[-0.200pt]{0.400pt}{3.132pt}}
\put(741.0,349.0){\rule[-0.200pt]{4.818pt}{0.400pt}}
\put(741.0,362.0){\rule[-0.200pt]{4.818pt}{0.400pt}}
\put(611,322){\circle{18}}
\put(751,218){\circle{18}}
\put(611.0,317.0){\rule[-0.200pt]{0.400pt}{2.409pt}}
\put(601.0,317.0){\rule[-0.200pt]{4.818pt}{0.400pt}}
\put(601.0,327.0){\rule[-0.200pt]{4.818pt}{0.400pt}}
\put(751.0,208.0){\rule[-0.200pt]{0.400pt}{4.818pt}}
\put(741.0,208.0){\rule[-0.200pt]{4.818pt}{0.400pt}}
\put(741.0,228.0){\rule[-0.200pt]{4.818pt}{0.400pt}}
\put(704,152){\makebox(0,0){$\star$}}
\put(704.0,151.0){\usebox{\plotpoint}}
\put(694.0,151.0){\rule[-0.200pt]{4.818pt}{0.400pt}}
\put(694.0,152.0){\rule[-0.200pt]{4.818pt}{0.400pt}}
\put(356,444){\raisebox{-.8pt}{\makebox(0,0){$\Diamond$}}}
\put(394,454){\raisebox{-.8pt}{\makebox(0,0){$\Diamond$}}}
\put(486,446){\raisebox{-.8pt}{\makebox(0,0){$\Diamond$}}}
\put(356.0,413.0){\rule[-0.200pt]{0.400pt}{14.936pt}}
\put(346.0,413.0){\rule[-0.200pt]{4.818pt}{0.400pt}}
\put(346.0,475.0){\rule[-0.200pt]{4.818pt}{0.400pt}}
\put(394.0,440.0){\rule[-0.200pt]{0.400pt}{6.986pt}}
\put(384.0,440.0){\rule[-0.200pt]{4.818pt}{0.400pt}}
\put(384.0,469.0){\rule[-0.200pt]{4.818pt}{0.400pt}}
\put(486.0,435.0){\rule[-0.200pt]{0.400pt}{5.059pt}}
\put(476.0,435.0){\rule[-0.200pt]{4.818pt}{0.400pt}}
\put(476.0,456.0){\rule[-0.200pt]{4.818pt}{0.400pt}}
\end{picture}
\end{center}
\vspace{-3mm}
\caption{
Quenched charmonium hyper-fine splitting for various tadpole improved
actions:
D234(${2\o 3}$) 3:1 ($\bullet$), SW 3:1 ($\circ$),
SW 1:1 ($\star$), FNAL ($\diamond$)~\protect\cite{FNAL} (a not fully 
relativistic formalism).
}
\label{fig:HFS}
\end{figure}

Data for  $c({\bf p})$ are given in table~1. They clearly demonstrate that
with TI rotational symmetry is restored to high accuracy for the D234 action.
Similar conclusions~\cite{LAT96} also hold for anisotropic lattices, even
for masses in the charmonium range. In table~2 we show results for $J$ and
$m_N/m_\rho$; in figure~1~\cite{LAT96} for the HFS. For $m_\rho/\sqrt{\si}$ 
we present the results~\cite{LAT95,SCRI,Urs} in words: Whereas the scaling 
violations
of the rho mass obtained with SW are much smaller (consistent with $\O(a^2)$)
than those with the Wilson action, the former are almost identical to that
of the isotropic D234 action. That they are almost identical is presumably
to some extent an accident; what is significant, is that these scaling
violations are almost 30\% on the coarsest lattice ($a\seq 0.4$~fm). For the
SW action this is not surprising, since other quantities obtained with this
action have similarly large errors.
It {\it is} surprising, though, for the D234 action, where  $c({\bf p})$,
$J$, and $m_N/m_\rho$ have errors of only a couple percent, 
even on the coarsest
lattice. Note that the D234 HFS, on the other hand, seems to have similarly
large errors as the rho mass on coarse lattices (much smaller scaling errors,
though, than the HFS of the SW action).

It is certainly very suggestive that  $m_\rho/\sqrt{\si}$ and the HFS
are exactly the quantities that depend most strongly, by far, on the value
of the clover coefficient. 
Also, L\"uscher et~al~\cite{LPCAC} have recently found, 
for the case of the SW action on Wilson glue, that the non-perturbative
clover coefficient is significantly larger than the tadpole estimate 
(which was used in the simulations described above), even on a relatively 
fine $a\seq 0.1$~fm lattice.

Although the data summarized above have some uncertainties that prevent 
them from being conclusive, it seems likely that the true clover coefficients 
of the (isotropic and anisotropic)  D234 actions are significantly larger 
than the values used above. With the correct values the D234 actions
can give accurate
results for all the indicators of scaling violations (a)--(e), already
on coarse lattices (certainly, the rho mass and HFS will increase, which
is what we want).

\section{Conclusions}

The use of the tadpole improved SW action on improved glue is a large
step forward compared to the use of the Wilson quark action. With the SW
action accurate (quenched) continuum extrapolations have been 
performed~\cite{SCRI,Urs}
{}from data in the range $a\seq 0.15-0.4$~fm. Further significant 
improvements
are possible with the D234 action. For both SW and D234 actions 
non-perturbative $\O(a)$ tuning~\cite{LPCAC} should be performed with
improved glue. The same applies to the determination of current
renormalization constants.\cite{LPCAC}
The methods of~\cite{LPCAC} can also be used on anisotropic lattices.
Tuning the additional coefficients might be significantly more complicated
in practice, 
but seems to be within reach of present technology. 
Actions and currents with no  $\O(a)$ and only small $\O(a^2)$ errors 
should give accurate results on coarse isotropic and anisotropic lattices.
Quenched QCD should essentially be ``solved'' within the next few years
(to the extent that it makes sense),
and there finally is hope for realistic simulations
of full QCD, heavy and heavy-light mesons, as well as glueballs and hybrids.


\section*{Acknowledgments}
I would like to thank my collaborators Mark Alford and Peter Lepage for
many discussions.
This work is supported by DOE grants DE-FG05-85ER25000 and DE-FG05-92ER40742.

\end{document}